\documentstyle[12pt]{article}
\renewcommand {\d}  {\hbox{d\hskip-1.1ex{\raise0.640ex\hbox{--}}
\skip 0.70ex}}
\newcommand   {\D}  {\hbox{D\hskip-1.9ex{\raise0.175ex\hbox{--}}
\hskip0.85ex}}

\begin{document}

\newcounter{buco}
\setcounter{equation}{0}
\setcounter{buco}{0}
\renewcommand{\thesection}{\Roman{section}}
\renewcommand{\theequation}{\arabic{section}\arabic{equation}}

% \setlength{\voffset}{-0.5in} \setlength{\textheight}{8.5in}
%\def\baselinestretch{2}
%\@addtoreset{equation}{section}
\def\theequation{\thesection.\arabic{equation}}
\def\theeqnarray{\thesection.\arabic{eqnarray}}

\begin{center}
\begin{Large}
\begin{bf}
 Covariant generalization of the ISGW quark model\\
\end{bf}
\end{Large}
\vspace{2.0cm}
D. Tadi\' c and S. \v Zganec \\
\vspace{1.5cm}
Physics Department, University of Zagreb, Bijeni\v cka c. 32, Zagreb,
Croatia
\end{center}
\vskip 2cm
\begin{center}
\bf Abstract
\end{center}
\baselineskip=24pt

\noindent
A fairly general Lorentz-covariant quark model of mesons is
constructed.
It has several versions whose nonrelativistic limit corresponds
to the well-known
Isgur, Scora, Grinstein, and Wise model. In the heavy-quark
limit, the covariant model
naturally and automatically produces the heavy-quark symmetry
results for meson decay constants
and semileptonic decay form factors. The meson decay constants
and the Isgur-Wise
functions are calculated for various versions of the covariant
model and compared
with other estimates. A general and adaptable structure of the
covariant model
ensures that it can be used to describe transitions involving
light and/or heavy mesons.

\newpage

\section{Introduction}
\vspace{2cm}

\indent

The well-established, simple, and often used, nonrelativistic
quark model of Isgur,
Scora, Grinstein, and Wise (ISGW) [1] has also been
employed [2,3,4,5]
in the investigations
of heavy-quark symmetry (HQS). Although the ISGW model
helped in HQS investigations , this
 nonrelativistic model was not capable [4,5] of properly reproducing
 all of the heavy-quark effective-theory (HQET)
relations among semileptonic meson decay form factors. It had to
be "relativized" to some extent [2,4,5]. Moreover, even
in the original paper [1] some compensation for relativistic
effects had to be
introduced with meson wave functions. In this way, useful
insights in the HQET
were gained and subleading corrections of
order $\overline{\Lambda} /m_{Q}$ were estimated [5]
(Here $\overline{\Lambda} \sim \Lambda_{QCD}$ and $m_{Q}$ is
the heavy-quark mass).

Thus it seemed useful to develop a fully covariant model that,
 in the nonrelativistic limit (NRL), goes into the
ISGW model. It turned out that such a covariant model
can, to a great extent, retain the simplicity, which was
an endearing and useful feature of the nonrelativistic model [1].

The covariant model can have a fairly general form [6] that
can be, if wanted,
specified in such a way as to lead to the ISGW model
in the NRL. The given
covariant formulation allows reasonable freedom in the
selection of model
parameters and model meson wave functions. They can
be selected to
reproduce a particular
Isgur-Wise function (IWF), what might provide a good
basis for the calculation of
$\overline{\Lambda}/ m_{Q}$ corrections.

An important feature in all variations of the proposed covariant
quark model (CQM) is the
description of valence quarks (antiquarks). They are parametrized
by the on-mass-shell
Dirac spinors, as it was the case in earlier models [1]
and in all subsequent
usages [2-6]. In a covariant model, such a description
might lead to
difficulties with the
covariant definition of a meson mass $M$. As shown in the next
section, this can be
resolved by introducing a scalar function that represents
the neutral sea with
 a momentum $K$ and vacuum  quantum numbers [6]. This
is an attractive feature,
as sea contributions must figure in a description of a hadron.
In the present
model, which takes into account only fluctuations
involving valence quarks (antiquarks),
the sea is described in the simplest possible way, as a
physical vacuum. The sea momentum function $F(K)$ has a
particularly simple form if one wants
to define a model that in the NRL goes into the ISGW model.

In the third section of this paper the meson decay constants and
the IWF are
calculated for this version of the CQM inspired by the ISGW
model. It turns
out that the covariant formulation takes care of the
relativistic effects, which previously had to be compensated
for by a phenomenological parameter $\kappa$ (see Fig. 1). The
sea function in this version
of the CQM is just a Dirac delta function, which ensures that a
meson has a properly defined on-mass-shell four-momentum $P$,
$(P^{2}=M^{2})$.

However, one could use a nontrivial $F(K)$ function in the
CQM. The form of such a function would influence the model
description of the physical quantities.
This is illustrated in the concluding sections of this paper
by calculating meson decay constants and the IWF for
a Gaussian $F(K)$.

\newpage

\section{Relativistic model(s) and the ISGW limit}
\vspace{2cm}
\setcounter{equation}{0}
\renewcommand{\theequation}{\arabic{section}.\arabic{equation}}

\indent

A meson $H$ with the four-momentum $P$ and  the mass
$M$ is covariantly represented by

\begin{displaymath}
|H(E,\vec{P}, M) \big> =
N \sum_{c,s_{1},s_{2}} \ \
\int  \ \big[ 4m_{Q}m_{d} \big] \ d^{4}p \
\delta^{4} (p^{2}-m^{2}_{Q}) \ \Theta (e)
\end{displaymath}

\begin{equation}
\cdot d^{4}q  \ \delta(q^{2}-m^{2}_{d})  \
\Theta (\epsilon)  \ d^{4}K  \ F(K)\
\delta^{(4)} (p+q+K-P)  \ \Theta (E)  \ \phi (l_{\perp})
\end{equation}

\begin{displaymath}
\cdot \overline{u}_{Q,s_{1}}^{c} (\vec{p})
\gamma_{5} v^{c}_{d,s_{2}} (\vec{q}) \ \
d^{+}_{d} (\vec{q}, c, s_{2}) b^{+}_{Q}
(\vec{p}, c, s_{1}) | 0 \big>
\end{displaymath}
Here, the index $d$ refers for concreteness, to a light
$d$ antiquark, whereas the index $Q$
denotes any of heavy quarks. The Dirac functions such as
$\delta (p^{2}-m^{2}_{Q})$,
combined with the corresponding step function $\Theta (e)$,
ensure that valence quarks
are on the mass shell. This is a characteristic feature of
the ISGW model. One can select, if desired, a
quark wave function $\phi (l_{\bot})$ which goes into the
ISGW wave function in the NRL.
Various momenta in $(2.1)$ are

\begin{displaymath}
l^{\mu}_{\bot} (P) = l^{\mu} - \frac{P^{\mu}(P \cdot l)}{M^{2}}
\end{displaymath}

\begin{equation}
l^{\mu} = \frac{1}{2} (p^{\mu} - q^{\mu})
\end{equation}

\begin{displaymath}
p^{\mu}= (e, \vec{p}) \ \  ; \ \ q^{\mu} = (\epsilon, \vec{q})
\end{displaymath}

\begin{displaymath}
l^{\mu}_{\bot}(M) = (0, \frac{\vec{p}-\vec{q}}{2})
\end{displaymath}
The ISGW limit is obtained if the wave function is selected as

\begin{equation}
\phi(l^{\mu}_{\bot}) = \frac{1}
{
\pi^{3/4} \beta^{3/2}_{S}
}
e^{
+(l^{\mu}_{\bot})^{2}/2 \beta^{2}_{S}
}
\end{equation}
Here $\beta_{S}$ corresponds to the variational solution [1]
with the
harmonic-oscillator wave functions.

The sea function $F(K)$ ensures that the meson mass M can be
covariantly defined.
Without $F(K)$, the Dirac delta function $\delta^{(4)}(p+ q - P)$
leads in the rest
frame $(\vec{P} = 0)$ to the momentum-dependent mass [1]
$M=(\vec{p}^{2}+m^{2}_{Q})^{1/2} + (\vec{p}^{2} + m^{2}_{d}) ^{1/2}$.
Generally, one has the same freedom in selecting $F(K)$
as one had in selecting
$\phi (l_{\bot})$. However, the ISGW state vectors of
the weak-binding limit
(WBL) [1] will be obtained if a simple form is selected:

\begin{equation}
F(K) = \delta^{(4)} \big[K^{\mu} - \frac{P^{\mu}}{M}
\big( \frac{P^{\nu}}{M}
\big(P-(p+q)\big)_{\nu}\big]
\end{equation}
In the meson rest frame:

\begin{equation}
K^{0}(\vec{P}=0)= [M-e-\epsilon]= \mu_{K} (\vec{p}, \vec{q})\ \ \
;\ \ \ \vec{K} (\vec{P}=0) = 0
\end{equation}
Obviously, as $\mu_{K}(\vec{p}, \vec{q})$ is not always
positive, the K does not
correspond to a physical, on-mass-shell particle. It can
be associated
with some sea contribution. This contribution can, in
principle, have a
less naive form than (2.4), which has been inspired by
the ISGW limit. For
example, one could try (see Sec. IV) the form

\begin{equation}
F(K) =
\delta^{(4)} \big[K^{\mu} - \frac{P^{\mu}}{M} \big( \frac{P^{\nu}}{M}
\big(P-(p+q)\big)_{\nu}\big]
\ e^{-\alpha K^{2}}
\end{equation}
It also leads to the ISGW model in the NRL.

After performing the integrations $d^{4}K$, $dp^{0}$,
and $dq^{0}$ in (2.1),
one is left with

\begin{displaymath}
|H (E, \vec{P}, M)> = N \sum_{c,s_{1},s_{2}} \int
d^{3}p \frac{m_{Q}}{e} d^{3}q \frac{m_{d}}{\epsilon}
\ \delta^{(4)}
[(p+q)^{\mu} - \frac{P^{\mu}}{M}
(\frac{E(e+ \epsilon)}{M} - \frac{\vec{P}(\vec{p}+\vec{q})}{M}) ]
\end{displaymath}

\begin{equation}
\cdot \phi (l_{\bot}) \ \overline{u}^{c}_{Q,s_{1}}
(\vec{p}) \gamma_{5} v^{c}_{d, s_{2}} (\vec{q}) \
d^{+}_{d} (\vec{q}, c, s_{2})\ b^{+}_{Q} (\vec{p}, c, s_{1}) |0>
\end{equation}
Using the notation

\begin{equation}
p^{\mu}_{\bot}= p^{\mu} - \frac{P^{\mu}}{M^{2}} (p \cdot P)\ \ \
; \ \ \ \ \ p^{\mu}_{\parallel}= \frac{P^{\mu}}{M^{2}} (p \cdot P)
\end{equation}
one realizes that the Dirac delta function in (2.7) constrains the
orthogonal components of the quark four-vectors, i.e.,

\begin{equation}
p^{\mu}_{\bot} + q^{\mu}_{\bot}= 0
\end{equation}
In the meson rest frame this gives the ISGW relation

\begin{equation}
\vec{p} + \vec{q} = 0
\end{equation}
By rewriting the complex $\delta$ function in (2.7) one
obtains a more
manageable form

\begin{displaymath}
|H(E,\vec{P},M)> = N \sum_{c,s_{1},s_{2}}
\int d^{3}p \frac{m_{Q}}{e}
d^{3} q \frac{m_{d}}{\epsilon}
\frac{1}{\frac{E^{2}}{M^{2}}(1-\frac{\vec{P}\vec{q}}{E \epsilon})}
\ \delta^{(3)}(\vec{q} + \vec{p} - \frac{\vec{P}}{M}
(p_{\parallel})T)
\end{displaymath}

\begin{equation}
\cdot  \phi (l_{\bot}) \
\overline{u}^{c}_{Q,s_{1}}
(\vec{p})
\gamma_{5} v^{c}_{d,s_{2}} (\vec{q})
\ d^{+}_{d} (\vec{q}, c, s_{2}) b^{+}_{Q} (\vec{p},c,s_{1})|0>
\end{equation}
Here the quantities $T$ and $p_{\parallel}$ are

\begin{equation}
T=1+ \frac{
\sqrt{m^{2}_{d}-m^{2}_{Q}+p^{2}_{\parallel}}
}
{p_{\parallel}}
\ \ \ ;\ \ \ \ \ p_{\parallel}= \frac{P \cdot p}{M}
\end{equation}

In the NRL and the WBL [1],

\begin{displaymath}
\vec{p}\,^{2} <<m_{Q}^{2} \ \ \ ;\ \ \  \vec{q}\,^{2}<<m_{d}^{2}
\end{displaymath}
\begin{displaymath}
E \rightarrow M \ \ \ \ ;\ \ \ \ e \rightarrow m_{Q}\ \ \ \
;\ \ \ \ \epsilon \rightarrow m_{d}
\end{displaymath}
\begin{equation}
 p_{\parallel} \rightarrow m_{Q} - \frac{\vec{P}\vec{p}}{M}\ \ \
\ \ \ \ \ ;
\ \ \  \vec{q} \rightarrow -\vec{p} + (m_{Q}+m_{d})
\frac{\vec{P}}{M}
\end{equation}
\begin{displaymath}
 (l^{\mu}_{\bot})^{2} \rightarrow 2m_{Q}
\frac{\vec{P}\vec{p}}{M} - \vec{p}\,^{2}
\ \ \ ;\ \ \
\overline{u}_{Q,s_{1}}(\vec{p}) \gamma_{5} v_{d, s_{2}}
(\vec{q}) \rightarrow \delta_{s_{1},-s_{2}}
\end{displaymath}
one finds

\begin{equation}
|H(E, \vec{P}, M)>_{NR,WB}=
N \sum_{c,s_{1},s_{2}} \int d^{3}p
\ [\frac{1}{
\pi^{3/4} \beta_{S}^{3/2}
}
e^{
(- \vec{p}\,
^{2}+2 m_{Q}
\frac{
\vec{P} \vec{p}
}{M})/_{2 \beta^{2}_{s}}}]
\end{equation}

\begin{displaymath}
\cdot \delta_{s_{1}-s_{2}} \  d^{+}_{d} (-\vec{p}+(m_{d}+m_{Q})
\frac{\vec{P}}{M}, c, s_{2})
b^{+}_{Q} (\vec{p}, c, s_{1}) |0>
\end{displaymath}
The substitution

\begin{equation}
\vec{p} = \vec{p'} + \frac{m_{Q}}{M} \vec{P}
\end{equation}
leads to the well-known [1-5] ISGW form

\begin{equation}
|H(E, \vec{P}, M)>_{NR,WB}=
N \sum_{c,s_{1},s_{2}} \int d^{3}p'
\ [\frac{1}{\pi^{3/4} \beta_{S}^{3/2}} e^{- \vec{p'}^{2}/2
\beta^{2}_{S}}]
\end{equation}

\begin{displaymath}
\cdot\delta_{s_{1},- s_{2}} \  d^{+}_{d} (- \vec{p'} +
\frac{m_{d}}{M} \vec{P}, c, s_{2}) b^{+}_{Q} (\vec{p'} +
\frac{m _{Q}}{M} \vec{P}, c, s_{1})|0>
\end{displaymath}

The full covariant forms (2.1) or (2.11) lead to fully
covariant predictions
for meson form factors in the CQM, as shown below. These states can
be covariantly normalized. With

\begin{equation}
<H(E', \vec{P'},M)| H (E, \vec{P},M)>= 2E \ \delta^{(3)}
(\vec{P} - \vec{P}')
\end{equation}
one finds for the state (2.7):

\begin{displaymath}
(2 \pi)^{6} 3 N(\vec{P})^{2} \int d^{3}p
\frac{m_{Q} m_{d}}{e \ \ \epsilon} [ \phi (l_{\bot})]^{2}
\frac{1}{
[\frac{E^{2}}{M^{2}}
(1-\frac{
\vec{P} \vec{q}
}{E \epsilon})]^{2}
}
\frac{1}{
(\frac{p_{\parallel}T}
{M})^{3} }
\end{displaymath}

\begin{equation}
\cdot \frac{1}{
[1+ \frac{
\vec{P} (e \vec{P}-E \vec{p})
}
{
EM \sqrt{m^{2}_{d}-m^{2}_{Q}+p^{2}_{\parallel}
}
}]}
\frac{p \cdot q + m_{Q} m_{d}}{m_{Q} m_{d}}
\delta^{(3)} ( \vec{P} - \vec{P}')|_{
\vec{q}= - \vec{p} + \frac{\vec{P}}{M} (p_{\parallel})T} =
2E \delta^{3} (\vec{P} - \vec{P}')
\end{equation}
 From this expression, $N(\vec{P})$ is

\begin{equation}
N(\vec{P}) = \frac{E}{M} N(0)
\end{equation}
where $N(0)$ can be calculated numerically.

\newpage

\section{Meson form factors and the Isgur-Wise function}
\vspace{2cm}
\setcounter{equation}{0}
\renewcommand{\theequation}{\arabic{section}.\arabic{equation}}

\indent

The form factors for $B \rightarrow D (D^{*})$ semileptonic
transitions are
defined in the standard way:

\begin{displaymath}
<D^{+} (\vec{P}_{D})| \overline{c} \gamma^{\mu} b |
\overline{B}^{0} (\vec{P}_{B})>=
\frac{1}{(2 \pi)^{3}}
[f_{+} (P_{B} +P_{D})^{\mu} + f_{-}(P_{B}-P_{D})^{\mu}]
\end{displaymath}

\begin{equation}
<D^{*+} (\epsilon,\ \vec{P}_{D})| \overline{c} \gamma^{\mu} b |
\overline{B}^{0} (\vec{P}_{B})>=
\frac{i}{(2 \pi)^{3}}
g \epsilon^{\mu \nu \rho \sigma} \epsilon^{*}_{\nu}
(P_{B}+P_{D})_{\rho} (P_{B}-P_{D})_{\sigma}
\end{equation}

\begin{displaymath}
<D^{*+} (\epsilon,\ \vec{P}_{D})| \overline{c}
\gamma^{\mu} \gamma_{5} b |
\overline{B}^{0} (\vec{P}_{B})>=
\frac{1}{(2 \pi)^{3}}
[f \epsilon^{* \mu}+ a_{+} (\epsilon^{*} \cdot P_{B})
(P_{B}+P_{D})^{\mu}
+ a_{-}(\epsilon^{*} \cdot P_{B})(P_{B}-P_{D})^{\mu}]
\end{displaymath}
Here the vector meson state $|D^{*}>$ is obtained from (2.1)
by replacing $\gamma_{5}$
by  $(\epsilon \gamma)$.
For example, one finds that

\begin{displaymath}
<D^{+}(\vec{P}_{D})| \overline{c} \gamma^{\mu} b | \overline{B^{0}}
(\vec{P}_{B})>= 3 (2\pi)^{3}  N_{D} (\vec{P}_{D}) N_{B}
(\vec{P}_{B}) \sum_{s_{1},s_{2},s'_{1}}
 \int d^{3}p' d^{3}p \ \frac{m_{c}  m_{b}}{e'  e}
\frac{m_{d}}{\epsilon '}
\end{displaymath}

\begin{equation}
\cdot
\frac{1}{
\frac{E^{2}_{D}}{M_{D}^{2}}
 (1- \frac{\vec{P}_{D} \vec{q'}}{E_{D} \epsilon '})}
\frac{1}{
\frac{E^{2}_{B}}{M_{B}^{2}}
 (1- \frac{\vec{P}_{B} \vec{q'}}{E_{B} \epsilon '})}
\delta^{(3)} [-\vec{p}+ \frac{\vec{P}_{B}}{M_{B}}
(p_{B \parallel}) T_{B} +
\vec{p'} - \frac{
\vec{P}_{D}
}
{M_{D}
}
(p_{D \parallel}) T_{D}]
\end{equation}

\begin{displaymath}
\cdot \phi_{D} (l_{D \bot}) \phi_{B} (l_{B\bot})[- \overline{
v}_{d,s_{2}}(\vec{q'})
 \gamma_{5} u_{c,s'_{1}} (\vec{p'})
\overline{u}_{c,s'_{1}} (\vec{p'}) \gamma^{\mu} u_{b, s_{1}}(\vec{p})
\overline{u}_{b,s_{1}} (\vec{p}) \gamma_{5} v_{d,s_{2}} (\vec{q'})]
\end{displaymath}
Here

\begin{displaymath}
p_{D \parallel} = \frac{E_{D}e' - \vec{P}_{D} \vec{p'}}{M_{D}}
\ \ \ \ \ \ \ \ \ \ \ \
p_{B \parallel} = \frac{E_{B}e - \vec{P}_{B} \vec{p}}{M_{B}}
\end{displaymath}

\begin{displaymath}
T_{D}= 1 + \frac{
\sqrt{m^{2}_{d}-m^{2}_{c} + (p_{D \parallel})^{2}}
}
{p_{D \parallel}} \ \ \ \ \ \ \ \ \ \
T_{B}= 1 + \frac{
\sqrt{m^{2}_{d}-m^{2}_{b} + (p_{B \parallel})^{2}}
}
{p_{B \parallel}}
\end{displaymath}

\begin{displaymath}
e'=\sqrt{m^{2}_{c} + \vec{p'}^{2}} \ \ \ \ \
\ \ \ \ \ e=\sqrt{m^{2}_{b} + \vec{p}\,^{2}}
\end{displaymath}

In the B-meson rest frame $(\vec{P}_{B}=0)$ this becomes

\begin{equation}
<D^{+}(\vec{P}_{D})| \overline{c} \gamma^{\mu} b|
\overline{B}^{0} (0)>=
3(2 \pi)^{3} N_{D} (\vec{P}_{D}) N_{B}(0)
\int d^{3}p'd^{3} p \frac{
m_{c} \ m_{b}
}{e' \ e} \frac{m_{d}} {\epsilon'}
\end{equation}

\begin{displaymath}
\cdot \frac{1}{
\frac{E_{D}^{2}}{M_{D}^{2}} (1-
 \frac{\vec{P}_{D} \vec{q'}}{E_{D} \epsilon '})
}
\delta^{(3)} (- \vec{p} + \vec{p'} -
\frac{\vec{P}_{D}}{M_{D}}
(p_{D \parallel})T_{D}) \phi_{D} \phi_{B} \cdot \{ V^{\mu} \}
\end{displaymath}

Here

\begin{displaymath}
\{ V^{\mu} \} = \frac{1}{8} Tr[ \gamma_{5}
(1+\frac{\not \! p'}{m_{c}} )\gamma^{\mu} (
1+\frac{\not \! p}{m_{b}} ) \gamma_{5}
(1-\frac{\not \! q'}{m_{d}} ) ]
\end{displaymath}

\begin{equation}
 = \frac{1}{2}
[\frac{q'^{\mu}}{m_{d}}+
\frac{p^{' \mu} (pq)}{m_{b}m_{c}m_{d}} -
\frac{q'^{\mu} (pp')}{m_{b}m_{c}m_{d}}+
\frac{p^{\mu} (p'q')}{m_{b}m_{c}m_{d}}+
\frac{p'^{\mu}} {m_{c}}+ \frac{p^{\mu}}{m_{b}} ]
\end{equation}

and

\begin{displaymath}
\vec{q'}= - \vec{p} = -  \vec{p'}+ \frac{
\vec{P}_{D}
}
{M_{D}} (p_{D \parallel}) T_{D}
\end{displaymath}

\begin{equation}
\phi_{B} = \frac{1}
{ \pi^{3/4} \beta^{3/2}_{B} }
exp [\frac{1}
{2 \beta^{2}_{B}}
(m^{2}_{b}- (p_{B  \parallel})^{2})]
= \frac{1}
{ \pi^{3/4} \beta^{3/2}_{B} }
exp [\frac{- \vec{p}\,^{2}}{2 \beta^{2}_{B}} ]
\end{equation}

\begin{displaymath}
\phi_{D}=
\frac{1}
{
\pi^{3/4} \beta^{3/2}_{D}
}
 exp [ \frac{1}
 {
 2 \beta^{2}_{D}
 }
 (m^{2}_{c} - (p_{D \parallel})^{2})]
\end{displaymath}
The trace which determines $\{ V^{\mu} \}$ is analogous to
formula (22)
of Ref. [4] and the $\not \! q' /m_{d}$ term can be connected
with the
momentum $ \tilde k$ of the same reference. That term contains
the Wigner
rotation of the light quark.
Comparison with (3.1) allows the extraction of the form factors,
 which
can be computed numerically. In order to check Lorentz covariance,
calculations leading to identical results have also been carried
out in the
D-meson rest frame $(\vec{P}_{D} = 0)$.

The decays with the vector meson $D^{*+}$ in the final states
are described
by expressions analogous to (3.3). The factor $ \{ V^{\mu} \} $
has to be
replaced as follows:

\begin{equation}
\{ V^{\mu} \} \rightarrow \{ V^{* \mu} \} \  or \  \{ A^{* \mu} \}
\end{equation}
Here

\begin{equation}
\{V^{* \mu}\}= Tr
[
\not \!  \epsilon ^{*}
\frac
{
\not \! p ' + m_{c}
}{
2m_{c}
} \gamma^{\mu}
\frac{
\not \! p + m_{b}
}{2m_{b}
}
\gamma_{5}
\frac{
\not \! q' - m_{d}
}{2m_{d}
} ]
\end{equation}

\begin{equation}
\{A^{* \mu}\}= Tr
[
\not \!  \epsilon ^{*}
\frac
{
\not \! p ' + m_{c}
}{
2m_{c}
} \gamma^{\mu} \gamma_{5}
\frac{
\not \! p + m_{b}
}{2m_{b}
}
\gamma_{5}
\frac{
\not \! q' - m_{d}
}{2m_{d}
} ]
\end{equation}

In the heavy-quark limit (HQL), one is tempted to identify
the heavy-quark
momenta with the heavy-meson momenta, for example,

\begin{equation}
\vec{p'} = \frac{m_{c}}{M_{D}} \vec{P}_{D}
\end{equation}
The Dirac delta-function constraints then determine

\begin{equation}
\vec{q'} = \frac{m_{d}}{M_{D}} \vec{P}_{D}
\end{equation}
This means that both valence quarks seem to travel as free particles.
Indeed, with (3.9), one finds that

\begin{equation}
m^{2}_{c} -(p_{D \parallel})^{2}=0 \ \ \ ;\ \ \
\phi_{D}= \frac{1}{\pi^{3/4} \beta^{3/2}}
\end{equation}
One has failed to account for the Wigner rotation of the
light quark [4]
and all information on the internal quark momenta is lost.
Thus, a more
reasonable choice is

\begin{equation}
\vec{p'} \rightarrow \frac{m_{c}}{M_{D}} \vec{P}_{D} + \vec{k'},
\ \ \ |\vec{k'}|/m_{c}<<1
\end{equation}
It leads to

\begin{displaymath}
\phi_{D_{HQL}}= \frac{1}{\pi^{3/4} \beta^{3/2}} exp
[\frac{1}{2 \beta^{2}} (m^{2}_{c} -(p_{D \parallel})^{2})]
\ \big{|}_{HQL}
\end{displaymath}

\begin{equation}
=\frac{1}{\pi^{3/4} \beta_{HQL}^{3/2}} exp
[-\frac{1}{2 \beta^{2}_{HQL}} (\vec{k'}^2 -\frac{
(\vec{P}_{D} \cdot \vec{k'})^{2}
}
{E_{D}^{2}
})]
\end{equation}

\begin{displaymath}
(p_{D \parallel})^{2}_{HQL}=
m^{2}_{c}+ \vec{k'}^{2} -
 \frac{
(\vec{P_{D}} \cdot \vec{k'})^{2}
}
{E_{D}^{2}} + O( \frac{1}{m_{c}})
\end{displaymath}
Furthermore, in the HQL, $m_{c} \rightarrow M_{D}$, so that

\begin{equation}
\vec{k'} \rightarrow \vec{p'} - \vec{P}_{D}
\end{equation}
An analogous procedure is carried out for the B meson.

If one had chosen (3.9) instead of (3.12), one would have obtained

\begin{equation}
\{ V^{\mu} \} = \frac{1}{4}
Tr[ \gamma_{5} (1+\not \! v' )\gamma^{\mu} (
1+\not \! v ) \gamma_{5} ]
\end{equation}

\begin{displaymath}
\not \! v'=\frac{\not \! p'} {m_{c}} \ \ \ ;
\ \ \ \not \! v =\frac{\not \! p} {m_{b}}
\end{displaymath}
Here the Wigner rotation of the light quark is absent.
 The expression (3.15) is
analogous to the expressions employed by Ref. [5]. However,
this reference
does keep some information on the internal quark momenta in the
valence-quark wave function, by retaining
some relativistic terms [5], and thus evades
the unacceptable result (3.11).

Finally, one finds, for example,

\begin{equation}
f_{+}(\vec{P}_{B}=0)_{HQL} \ =
 \ \int d^{3}p' I (p')
\frac{M_{B}+M_{D}}{4M_{B}M_{D}}
[1+ \frac{\tilde{\epsilon'}}{m_{d}} - \frac
{(\vec{P}_{D} \cdot \vec{\tilde{q'}})}
{(E_{D}+M_{D})m_{d}}]
\end{equation}

\begin{displaymath}
I(p') = 3(2 \pi)^{6} N_{D_{HQL}}(\vec{P}_{D})
N_{B_{HQL}}(0)
 (\frac{M_{D}}{E_{D}})^{2}
\frac{
m_{d}M_{D}
}{
E_{D} \tilde{\epsilon'} - \vec{P}_{D} \vec{\tilde {q'}}
}
\phi_{D_{HQL}} \phi_{B_{HQL}}
\end{displaymath}
Here we have used

\begin{displaymath}
\frac{e}{m_{b}} \rightarrow 1 \ \ \ \ \ \ ;
\ \ \ \ \ \ \ \frac{e'}{m_{c}} \rightarrow \frac{E_{D}}{M_{D}}
\end{displaymath}

\begin{displaymath}
\vec{g'} \rightarrow \vec{\tilde{q'}}\ \ ;\ \ \
\frac{\epsilon'}{m_{d}} \rightarrow
\frac{\tilde{\epsilon'}}{m_{d}} \ \ ;\ \ \
\tilde{\epsilon '}= \sqrt{
m^{2}_{d}+ \vec{\tilde{q'}}^{2}
}
\end{displaymath}

\begin{equation}
\vec{\tilde{q'}}= - \vec{k'} + \frac
{\vec{P}_{D}}{M_{D}
}
\sqrt{
m_{d}^{2}+(\vec{k'}^{2} -
\frac{(\vec{P}_{D} \vec{k'})^{2}}{E_{D}^{2}})}
\end{equation}

\begin{displaymath}
\frac{
\vec{g'}}{m_{b}} \rightarrow 0 \ \ ;\ \ \
\frac{\vec{p'}}{m_{c}} \rightarrow \frac{\vec{P_{D}}}{M_{D}} \ \ ;
\ \ \
\vec{k'} \rightarrow \vec{p'} - \vec{P_{D}}
\end{displaymath}

\begin{displaymath}
\phi_{B} \rightarrow \phi_{B _{HQL}}=
\frac{1}
{\pi^{3/4}\beta^{3/2}_{HQL}}
exp \big{[} \frac{
\vec{\tilde{q'}}^{2}
}
{2 \beta^{2}_{HQL}} \big{]}
\end{displaymath}

Similar expressions are readily obtained for other form factors.
Defining

\begin{equation}
F_{1}=f_{+}  \ \ ; \ \ \ A_{1} = \frac{1}{M_{B}+M_{D}}f \ \ ;
\ \ \ V=(M_{B}+M_{D}) g
\ \ ; \ \ \ A_{2}= - (M_{B}+M_{D}) a_{+}
\end{equation}
we find the well-known HQS relations [7]

\begin{equation}
F_{1} = V = A_{2} = \frac{1}{
1-\frac{Q^{2}}{(M_{B}+M_{D})^{2}}
}
A_{1}
\end{equation}
This immediately shows that the definition (3.12) has not
introduced any
 $\lambda/m_{Q}$ corrections. It only retained internal
quark momenta
 and the Wigner rotation, which is necessary if inconsistencies and
 contradictions are to be avoided [4]. The relations (3.19)
are valid only in
 the HQL. Then all form factors contain the same Isgur-Wise function
 (IWF), which is determined by

\begin{equation}
\xi (w) = Rf_{+} = RF_{1}
\end{equation}

\begin{displaymath}
R= \frac{2 \sqrt{M_{B} M_{D}}
}{
M_{B}+M_{D}}
\end{displaymath}
Here

\begin{displaymath}
w = v \cdot v' = \frac{(P_{D} \cdot P_{B})}{M_{D} M_{B}}
\end{displaymath}
In the B-meson rest frame ($\vec{P}_{B}=0$) one finds
\begin{equation}
\frac{
E_{D}
}{M_D} = w \ \ \ \ ;\ \ \ \ \ \
\frac{
|\vec{P}_{D}|
}
{M_{D}} = \sqrt{w^{2}-1}
\end{equation}
The expression (3.20) also satisfies the well-known [7]
constraint $\xi (1) = 1$ .
In Fig.1 our IWF is compared with that calculated by Amundson
[5], who obtained

\begin{equation}
\xi (w)_{A} = exp[ -
\frac{
m_{d}^{2}
}
{2 \kappa^{2} \beta^{2}_{HQL}
} (w-1)]
\end{equation}
Our curve (solid line in Fig.1) is calculated using the parameters
of Ref. [5]

\begin{equation}
\beta_{HQL}=0.42 \ GeV \ \ \ \ ;\ \ \ \ \ \ m_{d} = 0.33 \ GeV
\end{equation}

A meson decay constant $f_{H}$ is determined by the expression

\begin{displaymath}
\frac{1}{(2\pi)^{ \frac {3} {2}}} P^{\mu} f_{H} =
<0|: \overline{\psi}_{d}(0) \gamma^{\mu} \gamma_{5}
\psi_{Q} (0): |H (E, \vec {P},M)>
\end{displaymath}

\begin{equation}
=3 N(\vec{P}) \int d^{3}p
\frac{m_{Q}M}{eE}
\frac{M m_{d}}
{
E \epsilon - \vec{P} \vec{g}
}
 \phi(l_{\bot})
[\frac{
m_{d}p^{\mu}+m_{Q}q^{\mu}
}
{m_{d}m_{Q}}]
|_{\vec{q} = - \vec{p}+
\frac{
\vec{P}
}{M} (p_{\parallel})T}
\end{equation}
Here $\psi_{Q}(x)$ are valence quark fields [1]. The decay constant
$f_{H}$ is easily calculated in the frame $\vec{P}=0$. In order to
check covariance numerically, it has also been calculated for several
different $\vec{P}$ values. The results have always been identical.

With the parameters [1,5]

\begin{eqnarray}
\begin{array}{lll}
\beta_{S_{(u \overline{c})}} = 0.39 \ \  GeV
& \hspace{2cm} & m_{u} = m_{d} = 0.33 \ \  GeV\\
& \hspace{2cm} & m_{c} = 1.645 \ \  GeV\\
\beta_{S_{(u \overline{b})}} = 0.42 \ \  GeV
& \hspace{2cm} & m_{b} = 4.983 \ \  GeV\\
\end{array}
\end{eqnarray}
one finds that

\begin{equation}
f_{D}  = 258,8 \ MeV \ \ \ ;\ \ \
f_{B}  = 151,9 \ MeV
\end{equation}
In the HQL, the expression (3.24) takes the form

\begin{equation}
\frac{1}{(2\pi)^{ \frac {3} {2}}} P^{\mu} f_{H_{HQL}} =
3N_{HQL}(\vec{P})
\int d^{3}p (\frac{M}{E})^{2}
\frac{M m_{d}}
{
E \tilde{\epsilon}-\vec{P} \vec{\tilde {q}}
}
\phi_{HQL}(l_{\bot})
[\frac{
m_{d} P^{\mu}+ M \tilde{q}^{\mu}
}{M m_{d}}]
\end{equation}

In the HQL, one uses the average meson masses

\begin{equation}
\overline{M}_{D} =
\frac{
3M_{D^{*}} +M_{D}
}{4} = 1.975 \ GeV \ \ \ ;\ \ \
\overline{M}_{B} =
\frac{
3M_{B^{*}} +M_{B}
}{4} = 5.313 \ GeV
\end{equation}
and (3.23) in order to find the HQS result

\begin{equation}
f_{B_{HQL}}=
\sqrt{
\frac{
\overline{M}_{D}
}
{\overline{M}_{B}}
}
f_{D_{HQL}}
\end{equation}
The numerical values
\begin{equation}
f_{D_{HQL}}= 235,8 \ MeV \ \ \ \ ;\ \ \ \
f_{B_{HQL}}= 143,8 \ MeV
\end{equation}
which have been obtained using $(3.27)$, are quite close
to the result $(3.26)$,
showing that the model-determined corrections to the HQL
are about $5 \div 6 \%$.

\newpage

\section{Gaussian sea}
\vspace{2cm}
\setcounter{equation}{0}
\renewcommand{\theequation}{\arabic{section}.\arabic{equation}}

\indent

It might be useful to demonstrate the flexibility of
the expression (2.1) by
selecting a sea function F(K) that would be different
from the naive choice
(2.4). In principle, this could be based on some QCD
modelling of the sea
contribution. However, for illustrative purposes a
simple example can be
selected, which in the NRL and WBL goes into the ISGW state. Yet, it
leads to noticeably different results when used in the CQM. This is

\begin{equation}
F(K) = \delta^{(4)}
[K^{\mu} - \frac{P}{M}^{\mu}
(\frac{P}{M}^{\nu} (P-(p+q))_{\nu})]
\cdot
e^{-\alpha K^{2}}
\end{equation}
A simple arbitrary choice for the parameter $\alpha$ is

\begin{equation}
\alpha= 2^{-1} \beta^{-2}
\end{equation}
The integration over the sea momentum K gives

\begin{equation}
K^{\mu} = \frac{P^{\mu}}{M}
[M-
 \frac{
 E(e+ \epsilon) -
 \frac{
 \vec{P}^{2}
 }{M}
 (p_{\parallel})T
 }{M}]
 \end{equation}
 In the meson rest frame $(P^{\mu}=(M,\ 0))$, this goes into $(2.5)$.
  In the WBL, one finds that

 \begin{equation}
 \vec{K} =0 \ \ \  ; \ \ \  K^{0} \cong M- (m_{d}+m_{Q})
 \end{equation}
Thus the $K^{2}$, dependence has disappeared and (2.1)
in the NRL-WBL is again
the ISGW state (2.16). This conclusion is valid for any
meson frame, i.e., any $P^{\mu}$.

In the CQM, all manipulations are exactly analogous to
those presented
in the second section of this paper. In all formulas
one has to make the substitution

\begin{equation}
\phi (l_{\bot}) \rightarrow \phi_{K} =
\phi(l_{\bot}) e^{-\alpha K^{2}}
\end{equation}

In the HQL, one can use (4.3) and (3.12)-(3.17) in order to obtain

\begin{displaymath}
K^{2}=(M-p_{\parallel}-q_{\parallel})^{2} \rightarrow
\tilde{q}^{2}_{\parallel}
\end{displaymath}

\begin{equation}
(\phi_{K})_{HQL} \rightarrow \phi(l_{\bot})_{HQL}
e^{- \alpha \tilde{q}^{2}_{\parallel}}
\end{equation}
Again, all formal deductions (3.17) are repeated
with insertions of the factors
$exp[- \alpha \tilde{q}^{2}_{\parallel}]$ associated
with the product
 $\phi_{D_{HQL}} \cdot \phi_{B_{HQL}}$. In the
frame $\vec{P}_{B}=0$, for example, one has

\begin{equation}
\phi_{D_{HQL}} \phi_{B_{HQL}} \rightarrow \phi_{D_{HQL}}
\phi_{B_{HQL}}
\cdot exp \{ - \alpha [(\tilde{q}_{D \parallel})^{2} +
\tilde{\epsilon}^{2}] \}
\end{equation}

\newpage

\section{Numerical examples and discussion}
\vspace{2cm}
\setcounter{equation}{0}
\renewcommand{\theequation}{\arabic{section}.\arabic{equation}}

\indent

It is well known that various relativistic wave functions
(states) can lead
to the same state in the NRL. This has been illustrated in
Secs. II and
IV for two slightly different versions of the CQM. However,
the different versions
of the CQM lead to somewhat different
estimates of physical quantities. These differences persist
even in the HQL.

Owing to the relations (2.4), (4.1), or similar, the interplay
of the sea and the
valence-quark contribution leads to a meson mass M which is
not just a sum
of the valence-quark masses. However, the values of the quark masses
are interconnected with the quark wave function (2.3). For any mass
change, the variational procedure which leads to (2.3)
has to be repeated.
Or alternatively, the meson wave function $\phi$ can be determined
in some other model, for example, in a model inspired
by the Bethe-Salpeter equation.
For illustrative purposes, the mass $m_{d}$ has been
changed in the model
versions determined by (2.3) and (2.4) or (4.1). Such a
procedure, admittedly inconsistent,
has been used just to illustrate the flexibility of the CQM's.

\begin{center}
Table I
\end{center}

In Table I the meson decay form factors are calculated
using (2.4) and the
parameters (3.23), (3.25), and (3.28). The spectator
antiquark mass $m_{d}$
 has been arbitrarily changed, as discussed above. As
expected, the $f_{H}$ values
change with $m_{d}$,
but not dramatically, mostly by less than $20 \%$.
Deviations from the HQL relation (3.29) are more interesting.
They amount to abouth
 $9\%$ when the full CQM is employed. This indicates that,
in this version of
the model, the HQS relation (3.29) presents a very good
approximation.
Absolute values $f_{B}(CQM)$ are closer
(less than $5\%$ difference) to
$f_{B_{HQL}}$, as it should be expected with $m_{b}/m_{c}=3.3$.

\begin{center}
Table II
\end{center}

The results presented in Table II show that the sea
contributions can be very
important. Whereas the general pattern is similar to that displayed
in Table I, the absolute values of the meson decay
constants $f_{H}$ are much
smaller. For example, with $m_{d}=0.33GeV$,
$f_{D}(4.1)/f_{D}(2.4)=0.59$, and
$f_{B}(4.1)/f_{B}(2.4)=0.70.$ However, this has to be
taken more as an
illustration of the model flexibility than as a
serious prediction. The sea
descriptions (2.4) or (4.1) are very crude and one
should better refer to
them as to "mock sea" functions.

Other calculations of the meson decay form factors lead to a
broad range of values. A relativistic quark model with
centrally confined
quarks [8] gave values that were smaller up
to 50\%  ($f_{D}=130.6\ MeV$,
$f_{B}=90.9\ MeV$) than our (3.26) values or the
values in Table I. These values are
much closer to our values shown in Table II, which illustrates how
model-based predictions depend on details of the model construction.
The larger $f_{D}$ and $f_{B}$ values, as obtained
in the present CQM,
are closer to the results based on QCD sum rules, lattice
calculations,
and semilocal parton-hadron duality [9]. The estimates
in the QCD sum rules [10]
gave $f_{D}=120 \div 250\ MeV$ and $f_{B}=90 \div 200\ MeV$.
The predictions
of the lattice calculations [11] are in a similar
range $f_{D}=170 \div 230\ MeV$
 and $f_{B}=140 \div 220\ MeV$. Lower values,
$f_{D} \cong 80\ MeV$, $f_{B} \cong 130\ MeV$,
remarkably close to the values of Ref. [8], were
found in a potential quark model [12].

The IWF calculated in the CQM defined by (2.4) (see Fig. 1)
shows similar
behavior to the ISGW-Amudson [5] result obtained with the correction
factor $\kappa =0.6.$ Its slope $\rho$ defined by

\begin{equation}
\xi (w) = 1 - \rho^{2} (w-1) + O((w-1)^{2})
\end{equation}
is about $25 \%$  larger than the result of Refs. [5]. This
can be attributed to the fully relativistic character of the CQM,
including Wigner rotation. As shown in Ref. [4]  the Wigner rotation
increases the slope of IWF by about 20\%.

Figure 2 shows the influence of the "mock sea" contribution.
The solid
curve and the dotted curve correspond to the sea functions (2.4) and
(4.1), respectively. The use of (4.1) increases the slope $\rho$ by
$10 \%$. One is tempted to assume that the relativistic effects
(see Fig.1) might play a larger role in the calculation of
the IWF than
the sea effects. However, the model is too crude for
such far-reaching conclusions.

An arbitrary d-antiquark mass change, using $m_{d}=0.1$ GeV instead
of (3.23), produces virtually the same $\xi (w)$ curve with either
the (2.4) or the (4.1) sea description. The
slope $\rho=1.07$ is $9\%$ smaller
than the result based on (3.23) and (2.4) (see also Fig. 1).

A plethora of $\rho$ values can be found in the
literature. By fitting the data on
$B\rightarrow Dl\nu$ and using different ansatze
for the IWF [4,13-15], the
$\rho$ values have been found to be in the
range  $\rho=0.92 \div 1.57$. This
is slightly larger than the value based [2,3,16-18]
on ISGW model [1], $\rho \approx 0.8$.
Relativistic quark models [8,12] gave  $\rho=1.25$  and  $\rho=1.1$,
respectively. The QCD sum-rule estimates [19-26]
$\rho=1.0 \div 1.14$,
as well as the lattice computations [27-29]
$\rho=0.71 \div 1.35$, are more or
less in the same range as the estimates [4,13-15] based
on the $B\rightarrow Dl\nu$
data. The values of the slope parameter $\rho$ shown in
Figs. 1 and 2, which roughly span
the range $\rho=0.93 \div 1.29$, are similar to
fitting-data estimates
[4,13-15], or QCD sum-rule estimates [19-26], or
lattice [27-29] estimates. The value
obtained in the simplest CQM version of the ISGW model
is $\rho=1.17$.

Our results obtained using (2.4) are connected with the Close and
Wambach [4] deductions. In their case, the ISGW model
was relativized
sufficiently to make it covariant in the HQL. They have also taken
care to include the Wigner rotation of light quarks.
With $m_{Q}^{-1}$
corrections, their approach might also approximately, and adequatly,
describe lighter mesons (K).

When one starts with a covariant description, the HQL
follows automatically
by the  $m_{Q}^{-1}$ expansion. The HQS results are
readily obtained in this
limit. The Wigner rotation is also automatically
included in a covariant
procedure. Furthermore, the CQM can be used for the description
of the light mesons or the heavy-light
meson transitions. A very general structure of the model,
including the sea
function $F(K)$, provides for its great adaptability and
ability to model
various physical situations.

\newpage

\begin{center}
{\bf Table I.} Meson decay constants corresponding
to formula (2.4).
\end{center}

\vspace{1cm}

\begin{center}
\begin{tabular}{|c|c|c|c|c|}
\hline
$m_{d} (GeV)$ & $f_{D} (MeV)$ & $f_{D_{HQL}}(MeV)$ &
$f_{B} (MeV) $ & $f_{B_{HQL}} (MeV) $ \\ \hline
0.33 & 258.8 & 235.8 & 151.9 & 143.8 \\
0.3 & 251.4 & 233.2 & 149.3 & 142.2 \\
0.2 & 227.1 & 223.6 & 140.3 & 136.3 \\
0.1 & 204.3 & 212.6 & 131.2 & 129.6 \\
0.01 & 187.1 & 202.2 & 123.6 & 123.3 \\ \hline
\end{tabular}
\end{center}

\newpage

\begin{center}
{\bf Table II.} Meson decay constants corresponding
to formula (4.1).
\end{center}

\vspace{1cm}

\begin{center}
\begin{tabular}{|c|c|c|c|c|}
\hline
$m_{d} (GeV)$ & $f_{D} (MeV)$ & $f_{D_{HQL}}(MeV)$ &
$f_{B} (MeV) $ & $f_{B_{HQL}} (MeV) $ \\ \hline
0.33 & 157.4 & 146.2 & 105.7 & 89.1\\
0.3 & 153.8 & 144.5 & 104.4 & 88.1\\
0.2 & 141.0 & 137.8 & 99.6 & 84.0 \\
0.1 & 127.2 & 129.2 & 93.5 & 78.8 \\
0.01 & 114.5 & 120.5 & 87.1 &  73.5\\ \hline
\end{tabular}
\end{center}

\newpage

{\bf Figure captions}

\vspace{1cm}

{\bf Fig.1} The IWF's are shown as functions of $w$.
The dashed curve shows the IWF
calculated in Ref. [5] with $\kappa=1$. The dash-dotted
curve shows the result
of Ref. [5] with $\kappa=0.6$. The corresponding slope
parameter is $\rho \approx 0.93$.
The solid curve is obtained using our formula (3.20).
Its slope parameter
$\rho \approx 1.17$ corresponds to the ansatz (5.1).
All IWF's were calculated using the parameters (3.23).
%\overline{\hspace*{145 mm}}

{\bf Fig.2} Several IWF's are shown. The dashed curve,
with $\rho \approx 1.07$, is
obtained by using either (3.20) or (3.20) plus (4.7)
substitution with $m_{d}=0.1\ GeV$.
The solid curve, $\rho \approx 1.17$, is obtained using
(3.20) and $m_{d}=0.33\ GeV$,
the same as for the solid curve in Fig.1. The dotted
curve is obtained with the Gaussian sea
(4.7). Its slope parameter is $\rho \approx 1.29$. All
slope parameters
corresponds to the ansatz (5.1).

\end{document}